\documentclass[structabstract]{aa}
\usepackage{graphicx}
\usepackage{txfonts}
\usepackage[breaklinks=true]{hyperref}
\usepackage{natbib}

\usepackage{amsmath}

\newcommand\ensuretext[1]{\ensuremath{\text{#1}}}%
\newcommand\ergcms{\ensuretext{erg\,cm$^{-2}$\,s$^{-1}$}}%
\newcommand\phcms{\ensuretext{ph\,cm$^{-2}$\,s$^{-1}$}}%
\newcommand\phs{\ensuretext{ph\,s$^{-1}$}}%
\newcommand\g{\ensuremath{\gamma}}%
\newcommand\fermi{\textit{Fermi}/LAT}
\newcommand\rsn{\ensuremath{R_\mathrm{SN}}}%
\newcommand\mgas{\ensuremath{M_\mathrm{gas}}}%

\begin{document}
\title{Search for high-energy $\gamma$-ray emission from galaxies\\
  of the Local Group with \fermi}

   \titlerunning{Search for high-energy $\gamma$-ray emission from galaxies of the Local Group with \fermi}

   \author{J.-P. Lenain
     \inst{1,2}\fnmsep\thanks{Now at Landessternwarte, Universit\"at Heidelberg, K\"onigstuhl, 69117 Heidelberg, Germany}
     \and
     R. Walter
     \inst{1}
   }
   
   \offprints{J.-P.~Lenain\\
     \email{\href{mailto:jean-philippe.lenain@unige.ch}{jean-philippe.lenain@unige.ch}}}
   
   \institute{ISDC Data Centre for Astrophysics, Center for Astroparticle Physics, Observatory of Geneva, University of Geneva, Chemin d'Ecogia 16, CH-1290 Versoix, Switzerland
     \and
     LUTH, Observatoire de Paris, CNRS, Universit{\'e} Paris Diderot; 5 Place Jules Janssen, 92190 Meudon, France
   }

   \date{Received 20 June 2011 / Accepted 12 September 2011}

 
  \abstract
   {With the discovery of high-energy \g-ray emission from the Andromeda galaxy (M\,31) by the \fermi\ collaboration, normal galaxies begin to arise from the shadows for the first time, providing insight into cosmic ray acceleration in external galaxies.}
   {We search for high-energy \g-ray emission from those galaxies in the Local Group that have so far not been investigated: M\,81, M\,83, IC\,342, Maffei\,1, Maffei\,2, and M\,94.}
   {\fermi\ public data from August 4, 2008 to January 1, 2011 were analysed for these galaxies. We compared the results to other starburst and normal galaxies detected so far at high energies: the Magellanic clouds, M\,31, and the starburst galaxies M\,82 and NGC\,253.}
   {No significant detection is found in the data for the sources in our sample, and we derive upper limits on their photon flux. After comparing the results to other Local Group objects, we find that the derived upper limits are fully compatible with expectations from cosmic rays interacting with the interstellar medium within the host galaxies. In the case of M\,33 and M\,83, a detection in \fermi\ data should be imminent. The expected fluxes for the other sources in the sample are below the sensitivity of \fermi, even after 10 years of observation. Collective emission from compact objects in the host galaxies is also found to be negligible compared to the expected emission from cosmic ray interactions.
   }
   {}

   \keywords{gamma rays: galaxies --
     ISM: cosmic rays --
     radiation mechanisms: non-thermal
   }

   \maketitle
%

\section{Introduction}
\label{sec-intro}

The usual suspects for extragalactic high-energy \g-ray emission are active galactic nuclei (AGNs), and more particularly sources presenting powerful, relativistic jets pointing close to the line of sight, the so-called blazars, which are BL\,Lac objects or flat spectrum radio quasars. Their contribution to the extragalactic diffuse \g-ray emission as unresolved sources has been extensively discussed \citep[see e.g.][]{1996ApJ...464..600S,2007ApJ...659..958D}.
However, less powerful sources begin to be revealed as high-energy emitters. For instance, the starburst galaxies \object{M\,82} and \object{NGC\,253} have been detected at high energies with \fermi\ \citep{2010ApJ...709L.152A}. Recently, we have proposed the association of a previously unidentified \fermi\ source from the \fermi\ first source catalogue \citep[1FGL,][]{2010ApJS..188..405A} with the archetypal Seyfert\,2 galaxy \object{NGC\,1068} \citep{2010A+A...524A..72L}. We also discussed there the origin of the high-energy \g-ray emission of \object{NGC\,4945}, another Seyfert\,2 galaxy encompassing a starburst region in its core, and argued that in this case, the starburst activity within may be the dominant emitter with respect to the central AGN, while the AGN activity might play the main role at high energies in the case of NGC\,1068.

Normal galaxies also begin to arise from shadows in the GeV range, as can be seen by the recent discovery of \g-ray emission with \fermi\ from our neighbour, the Andromeda galaxy \citep[\object{M\,31},][]{2010A+A...523L...2A}. Normal and Seyfert galaxies have been thought to contribute to a small amount of the extragalactic diffuse \g-ray emission \citep[see e.g.][]{1976MNRAS.175P..23S}. However, recent studies have shown that their contribution can in fact be dominant compared to the one from blazars \citep[see e.g.][]{2010ApJ...722L.199F}. We present here a study of the search for high-energy \g-ray emission from the major galaxies pertaining to the Local Group.

We first describe the sample of sources and the method for the high-energy \g-ray data analysis in Section~\ref{sec-ana}, before discussing our results in Section~\ref{sec-disc} and drawing some conclusions in Section~\ref{sec-concl}.

\section{Observations and data analysis}
\label{sec-ana}

We focus our study on the major galaxies of the Local Group \citep[see e.g.][]{2005AJ....129..178K} for which no results at high energies have been discussed so far, that is \object{M\,81}, \object{M\,83}, \object{IC\,342}, \object{Maffei\,1}, \object{Maffei\,2}, and \object{M\,94}. The other galaxies from the Local Group for which high-energy emission has been reported are the small Magellanic cloud \citep[\object{SMC},][]{2010A+A...523A..46A}, the large Magellanic cloud \citep[\object{LMC},][]{2010A+A...512A...7A}, the Andromeda galaxy \citep[M\,31,][]{2010A+A...523L...2A}, M\,82 \citep{2010ApJ...709L.152A}, Cen\,A \citep[\object{NGC\,5128},][]{2010Sci...328..725F,2010ApJ...719.1433A}, and NGC\,253 \citep{2010ApJ...709L.152A}. Our primary goal is to search for signatures of high-energy \g-ray emission from interactions of cosmic rays accelerated by stellar processes (e.g. supernov\ae) with the interstellar medium in the host galaxy, as opposed to \g-ray emission from AGN activity.

For the different sources in our sample, the \fermi\ data analysis was performed using the publicly available data from August 4, 2008 to January 1, 2011. We used the unbinned likelihood method \citep{2009ApJ...697.1071A} from the public \textit{Science Tools} (version \texttt{v9r18p6}) analysis software provided by the \textit{Fermi} collaboration. The data analysis was carried out following the procedure recommended by the \fermi\ collaboration\footnote{see \href{http://fermi.gsfc.nasa.gov/ssc/data/analysis}{http://fermi.gsfc.nasa.gov/ssc/data/analysis}.}, using diffuse class events, with the \texttt{P6\_V3} instrument responses, in the energy range of 200\,MeV--200\,GeV. The Galactic diffuse emission was modelled using the public model \texttt{gll\_iem\_v02}, while the isotropic model \texttt{isotropic\_iem\_v02} was used to account for both extragalactic, unresolved diffuse emission and residual instrumental background. These two models are available as part of the public \textit{Science Tools}.
For each source in the sample, we selected events from a circular region of 10\degr\ of radius around the nominal position of the galaxy of interest. Given the distances of the sources considered for our analysis -- the closest one being Maffei\,2 lying at 2.8\,Mpc -- the objects are assumed to be point-like at high energy. Indeed, the apparent dimensions of the host galaxies are mostly below 10\arcmin, i.e. smaller than the point spread function of the \fermi\ in the major part of the considered energy range.

In the modelled reconstruction of the sources, all the objects included in the 1FGL catalogue \citep{2010ApJS..188..405A} are accounted for. Moreover, other sources, for which test statistics \citep[TS, see e.g.][]{1996ApJ...461..396M} are above 25 -- roughly equivalent to a 5$\sigma$ detection -- and which are not reported in the 1FGL catalogue, are also included in the source models. These potential source candidates were selected from the \fermi\ count maps above 1\,GeV, where the \fermi\ point spread function is below $\sim$0.5\degr. For each object added in this way to the model, as well as for the source of interest, we first run a likelihood test, using \texttt{gtlike}, to assess whether the new introduced source is significant. If the corresponding TS is above 25, revealing a detection, the position of the new source is optimised using the tool \texttt{gtfindsrc}, whereas for the sources included in the 1FGL catalogue, we fixed their position to the values from the catalogue. The spectral parameters of the additional source are then refined on the corresponding best-fit position, using \texttt{gtlike}.

\begin{table*}
  \caption{Properties of galaxies from the Local Group.}
  \label{tab-ana_results}
  \centering
  \begin{tabular}{lccccc}
    \hline\hline
    Source         & d                   & \rsn                                    & \mgas                       & $L_\g$                                     & TS\\
                   & kpc                 & yr$^{-1}$                                & $10^9 M_{\sun}$               & $10^{43}$\,\phs                            & \\
    \hline
    Milky Way      & ...                 & $0.02 \pm 0.01^\mathrm{(1)}$              & $6.5 \pm 2.0^\mathrm{(2)}$    & $0.12 \pm 0.03^\mathrm{(3)}$                & ...\\
    LMC            & 50$^\mathrm{(4)}$     & $0.005 \pm 0.002^\mathrm{(5)}$            & $0.67 \pm 0.08^\mathrm{(6)}$  & $(7.8 \pm 0.8) \times 10^{-3 \,\mathrm{(7)}}$  & 1110.7$^\mathrm{(7)}$\\
    SMC            & 61$^\mathrm{(8)}$     & $(1 \pm 0.2) \times 10^{-3 \,\mathrm{(9)}}$  & $0.70 \pm 0.06^\mathrm{(10)}$ & $(1.6 \pm 0.4) \times 10^{-3 \,\mathrm{(11)}}$ & 138.8$^\mathrm{(11)}$\\
    M\,31          & 780$^\mathrm{(12)}$   & $0.045 \pm 0.015$                       & $7.66 \pm 2.38^\mathrm{(13)}$ & $(6.6 \pm 1.4) \times 10^{-2 \,\mathrm{(14)}}$ & 28.9$^\mathrm{(14)}$\\
    \object{M\,33} & 847$^\mathrm{(15)}$   & $0.050 \pm 0.015$                       & $2.23 \pm 0.84^\mathrm{(16)}$ & $< 5.0 \times 10^{-2 \,\mathrm{(14)}}$         & $<25^\mathrm{(14)}$\\
    Maffei\,2      & 2800$^\mathrm{(17)}$  & ?                                       & ?                            & $< 1.66$                                 & 0.0\\
    Maffei\,1      & 3010$^\mathrm{(18)}$  & ?                                       & ?                            & $< 3.03$                                 & 2.5\\
    IC\,342        & 3280$^\mathrm{(19)}$  & $0.18 \pm 0.10^\mathrm{(20)}$             & $4.0 \pm 0.8^\mathrm{(21)}$   & $< 6.40$                                  & 1.8\\
    M\,82          & 3530$^\mathrm{(22)}$  & $0.2 \pm 0.1^\mathrm{(1)}$                & $2.5 \pm 0.7^\mathrm{(1)}$    & $2.39 \pm 0.75^\mathrm{(1)}$                & 46.2$^\mathrm{(1)}$\\
    NGC\,4945      & 3600$^\mathrm{(23)}$  & $0.3 \pm 0.2^\mathrm{(24)}$               & $4.2^\mathrm{(25)}$           & $4.09 \pm 0.92^\mathrm{(26)}$               & 85.3$^\mathrm{(26)}$\\
    M\,81          & 3630$^\mathrm{(27)}$  & $0.008 \pm 0.002^\mathrm{(28)}$           & $5.16 \pm 1.72^\mathrm{(29)}$ & $< 1.55$                                  & 0.0\\
    NGC\,253       & 3940$^\mathrm{(30)}$  & $0.2 \pm 0.1^\mathrm{(1)}$                & $2.5 \pm 0.6^\mathrm{(1)}$    & $1.12 \pm 0.78^\mathrm{(1)}$                & 23.0$^\mathrm{(1)}$\\
    M\,83          & 4500$^\mathrm{(23)}$  & $0.050 \pm 0.025^\mathrm{(31)}$           & $5.5 \pm 1.1^\mathrm{(31)}$   & $< 1.33$                                  & 5.8\\
    M\,94          & 4660$^\mathrm{(32)}$  & $0.04 \pm 0.02^\mathrm{(33)}$             & $0.56 \pm 0.11^\mathrm{(33)}$ & $< 3.19$                                  & 0.0\\
    NGC\,1068      & 14000$^\mathrm{(34)}$ & $0.20 \pm 0.08^\mathrm{(35)}$             & $4.4^\mathrm{(36)}$           & $33.8 \pm 8.9^\mathrm{(26)}$                & 68.6$^\mathrm{(26)}$\\
    \hline
  \end{tabular}
  \tablebib{
    (1)~\citet{2010ApJ...709L.152A}; (2) \citet{1993AIPC..278..267D}; (3) \citet{2010ApJ...722L..58S}; (4) \citet{2009ApJ...697..862P};
    (5) \citet{1994ApJS...92..487T}; (6) \citet{2003MNRAS.339...87S,2008ApJS..178...56F}; (7) \citet{2010A+A...512A...7A}; (8) \citet{2005MNRAS.357..304H};
    (9) \citet{1981ApJ...243..736S}; (10) \citet{1999MNRAS.302..417S,2007ApJ...658.1027L}; (11) \citet{2010A+A...523A..46A}; (12) \citet{1998ApJ...503L.131S};
    (13) \citet{2006A+A...453..459N,2009ApJ...695..937B}; (14) \citet{2010A+A...523L...2A}; (15) \citet{2004A+A...423..925G}; (16) \citet{2010A+A...522A...3G};
    (17) \citet{2003A+A...408..111K}; (18) \citet{2003ApJ...587..672F}; (19) \citet{2004AJ....127.2031K}; (20) \citet{2005PhRvL..95q1101A}; (21) \citet{1979A+A....80..255R};
    (22) \citet{2002A+A...383..125K}; (23) \citet{2002A+A...385...21K}; (24) \citet{2009AJ....137..537L}; (25) \citet{2008A+A...490...77W}; (26) \citet{2010A+A...524A..72L};
    (27) \citet{2007AJ....133.1883K}; (28) \citet{1997ApJS..112...49M}; (29) \citet{2010AJ....140..663G}; (30) \citet{2003A+A...404...93K}; (31) \citet{2004A+A...422..865L};
    (32) \citet{2003A+A...398..467K}; (33) \citet{2011AJ....141...23B}; (34) \citet{2007ApJS..173..185G}; (35) \citet{1982ApJ...263..576W,1994ApJ...421...92B,2003A+A...401..519M}; (36) \citet{1990ApJ...351..422S}.
  }
\end{table*}

For all the sources of interest in the sample, the corresponding TS are found to be below 25. We thus computed upper limits at 2$\sigma$ confidence level on their flux in the 200\,MeV--200\,GeV energy range. No specific energy spectra are assumed to derive these upper limits, the photon indices of the sources of interest are left free to vary in the models. Table~\ref{tab-ana_results} summarises the results of our analysis on the source sample, as well as results for other galaxies studied in the \textit{Fermi} era. Following are some notes concerning peculiar sources.

\begin{figure}
  \centering
  \includegraphics[width=0.9\columnwidth]{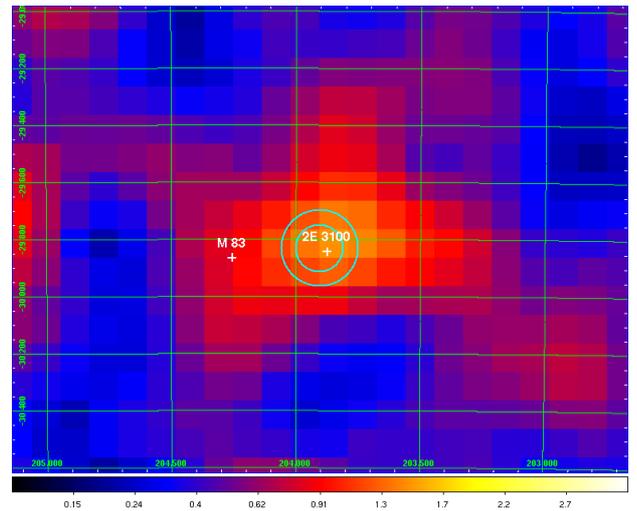}
  \caption{Smoothed count map of the region of interest around M\,83, in the 1\,GeV--200\,GeV energy range. The nominal positions of M\,83 and the blazar 2E\,3100 are shown in white. The best-fit position, at 68\% and 95\% confidence levels, of the \fermi\ excess is shown in blue.}
  \label{fig-M83_map}
\end{figure}

\subsection*{M\,81}

For the search of \g-ray emission from M\,81, the strong starburst emitter M\,82 is located 0.62\degr\ away from M\,81, hence contaminating the region around M\,81, given that the point spread function of the LAT degrades strongly at low energies, which prevents us from deriving constraining upper limits on the flux of M\,81. The corresponding upper limit given in Table~\ref{tab-ana_results} should thus be considered as conservative.

\subsection*{M\,83}

A significant signal is found in the vicinity of M\,83, with a TS=30.1 (5.5$\sigma$). However, the best-fit position of the excess turns out to be compatible with the nominal position of a blazar, 2E\,3100, lying at 0.33\degr\ from the position of M\,83. The \fermi\ source is not spatially compatible with M\,83, as can be seen in the count map shown in Fig.~\ref{fig-M83_map} for the 1\,GeV--200\,GeV energy range.

\subsection*{IC\,342}

IC\,342 lies at 10.6\degr\ from the Galactic plane. Given the proximity of the Galactic plane for this source lying at $b=+10.6\deg$, we extracted the events from a region of 10\degr\ of radius centred on a position displaced by 5.7\degr\ towards north-east compared to the nominal position of IC\,342, in order to reduce the fraction of the emission from the Galactic background in the region of interest. For this analysis, four sources were added to the source model in addition to those from the 1FGL catalogue, and the signal extracted around IC\,342 amounts to 1.3$\sigma$.

\subsection*{Maffei group}

The two galaxies Maffei\,1 and Maffei\,2 are separated by only 0.68\degr, making it difficult to derive high-energy properties for these two objects separately with \fermi, given the angular resolution of the LAT\footnote{see \href{http://www-glast.slac.stanford.edu/software/IS/glast_lat_performance.htm}{http://www-glast.slac.stanford.edu/software/IS/glast\_lat\_performance.htm}}. Moreover, they are located at less than 1\degr\ from the Galactic plane. For this reason, no precise constraints on their supernova rate or the total gas mass were found in the literature.

\subsection*{M\,94}

M\,94 is a spiral galaxy viewed almost face-on. The closest source from the 1FGL catalogue lies at $\sim$4\degr\ from M\,94, easing any analysis of this region. The signal extraction from M\,94 amounts to 0.0$\sigma$.

\section{Discussion}
\label{sec-disc}

High-energy \g-ray emission is expected from starburst galaxies, as well as normal galaxies, through cosmic ray interactions with the interstellar medium. Cosmic rays, which are mainly made up of charged, light hadrons and leptons that are accelerated in these galaxies, interact with the ambient interstellar medium, creating pions which will in turn decay into \g-rays for the hadrons, and bremsstrahlung and inverse Compton radiations for the leptons \citep[see e.g.][]{2010ApJ...722L..58S}. Based on a simple model, \citet{2001ApJ...558...63P} suggest there is a relationship between the expected \g-ray luminosity, the supernova rate \rsn, and the total gas mass \mgas\ for a given galaxy:

\begin{equation}
  \begin{aligned}
  F_\g (>100\,\mathrm{MeV}) = & 2.34 \times 10^{-8} \left(\frac{\rsn}{\rsn^\mathrm{MW}}\right) \left(\frac{\mgas}{10^8 M_{\sun}}\right)\\
   & \times \left(\frac{d}{100\,\mathrm{kpc}}\right)^{-2}\,\phcms
   \end{aligned}
\end{equation}
where \ensuremath{\rsn^\mathrm{MW}} is the supernova rate in the Milky Way and $d$ the distance of the considered galaxy.

\begin{figure}
  \centering
  \includegraphics[width=\columnwidth]{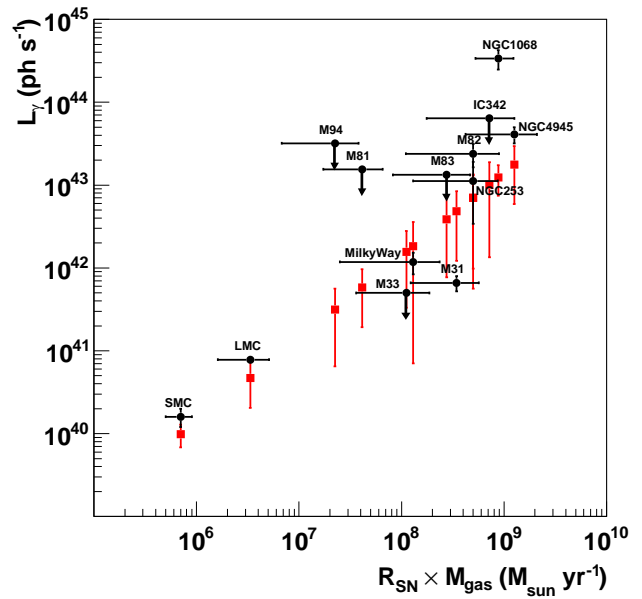}
  \caption{\g-ray luminosity, or the upper limit at 2$\sigma$ confidence level on the luminosity, of the different sources in the sample and other known high-energy emitting starburst and normal galaxies, given against the supernova rate times the total gas mass in these objects. The red square points show the expectations on the luminosity from the model of \citet{2001ApJ...558...63P}, accounting for the uncertainties on \rsn\ and \mgas.}
  \label{fig-relation}
\end{figure}

Figure~\ref{fig-relation} shows the \g-ray luminosity measured with \fermi\ for the different known high-energy emitting starburst and normal galaxies, against $\rsn \times \mgas$, and the upper limits at 2$\sigma$ confidence level on the fluxes derived from the sources in our sample. For comparison, we added the Seyfert\,2 galaxy NGC\,1068, also detected at high energies \citep{2010A+A...524A..72L}. This object encompasses a starburst region in its centre as well, but we argued in \citet{2010A+A...524A..72L} that the associated high-energy \g-ray emission is more likely due to the central AGN activity than from the starburst activity only, owing to its relatively high \g-ray luminosity.

As can be seen in Fig.~\ref{fig-relation}, the upper limit reported by the \fermi\ collaboration on M\,33 \citep{2010A+A...523L...2A} is very stringent compared to expectation, if the high-energy flux is indeed directly connected to $\rsn \times \mgas$, that is, if the high energy emission is solely due to cosmic ray interactions with the interstellar medium. In such a case, one could very soon expect a detection of this source with the \fermi. A non-detection would mean a different cosmic ray content in this galaxy, e.g. a lower supernov\ae\ rate than previously thought or a particularly efficient electron escape from the host galaxy to prevent high-energy emission. It is also noticeable that the current upper limit on M\,83 reported here also shows that a positive detection could be imminent. Apart from the case of NGC\,1068, where the high-energy \g-ray emission could be dominated by the central AGN \citep[see][]{2010A+A...524A..72L}, the observations are fully consistent with the expectations from the model of \citet{2001ApJ...558...63P}.

An alternative expectation of \g-ray emission from these sources resides in the recent discovery by \citet{2010ApJ...717..825D} of giant \g-ray bubbles in the inner Milky Way, whose origin is still unclear. \citet{2010ApJ...724.1044S} extensively discuss different possible mechanisms for the origin of this large-scale \g-ray emission. This emission more likely comes from inverse Compton scattering of a hard population of cosmic ray electrons, and may have the same origin as the microwave haze seen in \textit{WMAP} data \citep{2008ApJ...680.1222D}. If such high-energy emission is common in normal galaxies, one could also expect to detect that signal in the galaxies of the Local Group, in addition to the emission discussed above. In this case, the high-energy \g-ray flux of these objects is expected to be larger than the flux predicted in the model by \citet{2001ApJ...558...63P}. In the particular case of M\,33, such emission can already be ruled out given the constraining upper limit put on its \g-ray flux.

High-energy \g-ray emission could also arise from the collective emission of galactic sources within the host galaxies, such as pulsars or pulsar wind nebul\ae. To evaluate the contribution of such emission, we used the luminosities of pulsars and the few pulsar wind nebul\ae\ detected by \fermi\ \citep{2010ApJS..187..460A,2011ApJ...726...35A}, in order to estimate the expected high-energy flux from the collective emission, assuming that these Galactic objects are located in a hypothetical galaxy located at 3\,Mpc, which is about the mean distance for the source in our sample. We intentionally over-estimate the predicted flux by assuming 1000 of such sources in the host galaxy, which is much more numerous than what is currently detected in our own Galaxy\footnote{46 significant pulsars are reported in \citet{2010ApJS..187..460A}, and 10 pulsar wind nebul\ae\ are discussed in \citet{2011ApJ...726...35A}.}. Even under these highly optimistic conditions, the overall contribution from pulsars is expected to be $F(E>100\,\mathrm{MeV}) \sim 5 \times 10^{-14}$\,\ergcms, and the one for the pulsar wind nebul\ae\ amounts to $F(E>100\,\mathrm{MeV}) \sim 10^{-13}$\,\ergcms. Such estimations are well below the sensitivity performances of the \fermi\ within two years of observations\footnote{see \href{http://www-glast.slac.stanford.edu/software/IS/glast_lat_performance.htm}{http://www-glast.slac.stanford.edu/software/IS/glast\_lat\_performance.htm}.}. The total contribution of galactic objects in the sources of our sample is thus expected to be negligible compared to the global emission expected from cosmic ray interactions with the local interstellar medium.

\section{Conclusions}
\label{sec-concl}

We have searched for high-energy \g-ray emission from galaxies belonging to the Local Group. Apart from the Andromeda galaxy (M\,31), already established as a source of \g-ray emission by the \fermi\ collaboration \citep{2010A+A...523L...2A}, and NGC\,4945 \citep{2010A+A...524A..72L}, all the other sources have not yet been detected at high energy. We derived upper limits on their photon flux, and found that these limits are fully compatible with the expectations from the model of \citet{2001ApJ...558...63P} for interactions of cosmic rays with the local interstellar medium. In the case of M\,33 and M\,83, a detection with \fermi\ should be imminent, if this model holds. The expected fluxes for the other sources in the sample are below the sensitivity of \fermi, even after ten years of observation. This is consistent with the work of \citet{2010ApJ...722L.199F}, who estimate that there are around five normal galaxies that are individually detectable with \fermi. A contribution from the collective emission of compact objects in the host galaxies is found to be negligible compared to such expected emission.

Deeper observations provided by the \fermi\ all sky survey strategy will possibly allow for the detection of other normal, non-active galaxies in the near future.

\begin{acknowledgements}
We thank the anonymous referee, whose comments helped in improving the quality of this manuscript.

This research has made use of data and software provided by the Fermi Science Support Center, managed by the HEASARC at the Goddard Space Flight Center. This research also made use of NASA's Astrophysics Data System (ADS), of the SIMBAD database, operated at the CDS, Strasbourg, France, and of the NASA/IPAC Extragalactic Database (NED) which is operated by the Jet Propulsion Laboratory, California Institute of Technology, under contract with the National Aeronautics and Space Administration.
\end{acknowledgements}

\bibliographystyle{aa}  
\bibliography{Fermi_NormalGalaxies.bbl}

\end{document}